\documentclass[%
 reprint,
 superscriptaddress,
 amsmath,amssymb,
 aps,
]{revtex4-1}

\usepackage{graphicx}
\usepackage{dcolumn}
\usepackage{bm}
\usepackage{xcolor}
\usepackage{float}
\usepackage{txfonts}
\usepackage{multirow}
\begin{document}

\title{Non-negligible influence of shape inheritance and staggering on $\alpha$ decay}

\author{Ruixiong Li}
\affiliation{School of Nuclear Science and Technology, Lanzhou University, 730000 Lanzhou, P.R. China}
\author{Jingyu Xiao}
\affiliation{School of Nuclear Science and Technology, Lanzhou University, 730000 Lanzhou, P.R. China}
\author{Hongfei Zhang}
\affiliation{School of Physics, Xi'an Jiaotong University, 710049 Xi'an, People's Republic of China}
\affiliation{Joint Department for Nuclear Physics, Institute of Modern Physics, Chinese Academy of Sciences, and Lanzhou University, 730000 Lanzhou, People's Republic of China}
\author{Nana Ma}\email[]{mann@lzu.edu.cn}
\affiliation{School of Nuclear Science and Technology, Lanzhou University, 730000 Lanzhou, P.R. China}
\affiliation{Frontiers Science Center for Rare Isotopes, Lanzhou University, Lanzhou 730000, People's Republic of China
	Isotopes}
\begin{abstract}

A series of findings in machine learning (ML) and decay theory are captured while exploring the role of deformation and preformation factors in $\alpha$ decay. We provide a novel and practical paradigm for developing physics-driven machine learning in nuclear physics research by introducing known decay theory and statistical correlation analysis. Furthermore, this analysis verifies the Geiger-Nuttall law, and the relationship between the decay energy and $\alpha$ formation amplitude, also releases a signal that nuclei with hexadecapole deformation are more likely to form $\alpha$ clusters. In particular, we identify two novel phenomena, \textit{shape inheritance}, in which the deformation properties are partially transmitted from parent to daughter nuclei; and half-life \textit{inversion} due to shape staggering of adjacent even-even nuclei. This phenomenon occurs frequently in neutron-deficient nuclei near lead isotopes, which is consistent with shape coexistence in experiments. Surprisingly, it reappeared within the predicted half-life of the 119 and 120 isotope chains in the eighth period of the periodic table. The half-life considering the inversion effect is preferable for the study of new nuclides and shape coexistence in experiments.

\end{abstract}

\pacs{23.60.+e, 23.70.+j, 07.05.Mh}

\maketitle


\textit{Introduction.}-
The two-step mechanism of $\alpha$ decay provides a practical description, which involves the formation of $\alpha$ clusters on the surface of the decaying nucleus, followed by quantum penetration of the Coulomb barrier and the centrifugal barrier \cite{Gamow,GC}. It has dominated convincing descriptions of the main decay mode in nature and is especially indispensable for the experimental identification of new nuclides and nuclear structures \cite{Oganessian}. Therefore, factors concerning the formation and emission of $\alpha$ particles, the potential barriers, and penetration probabilities affect the precision of the description of $\alpha$ decay to varying degrees \cite{qichong1}.

The probability of $\alpha$ particle being preformed in a parent nucleus is quantified by the preformation factor \cite{preformation,qichong2,zhang}. It encapsulates the microscopic dynamics of nucleons aggregating into a $\alpha$ cluster. In decay theory, a characteristic closely related to the $\alpha$ formation amplitude is nuclear deformation \cite{qichong1,Andreyev1,Andreyev2}, which is quantified by deformation parameters such as quadrupole deformation ($\beta_{2}$) and hexadecapole deformation ($\beta_{4}$) \cite{Xu}. In addition, a series of studies that focus on the influence of deformation on various decay modes \cite{deformation1,deformation2,Guo}, which alter the geometric distribution of potential barriers, especially the anisotropy of the height and width of the barrier, also emphasize the indispensibility of deformation. 

The complex deformation and shape coexistence of nuclei in the nuclear region around lead isotopes \cite{Heyde,Andreyev3,May}, especially neutron-deficient nuclei, where nuclei exhibit spherical, oblate and prolate deformation, and rapid transitions between them, were reported in reliable experimental and theoretical work. These findings will trigger new research on the effects of deformation and preformation factors on $\alpha$ decay, especially for the synthesis of elements in the eighth period.

Machine learning (ML), a data-driven approach to predictive modeling \cite{ML1}, has emerged as a transformative tool in nuclear physics. ML captures nonlinear patterns and relationships within large and disordered datasets. This novel perspective offers promising opportunities to identify hidden complexities, achieve precise predictions of physical phenomena, and improve forecasts \cite{ML2}. This inspires researchers to consider the decay of nuclei with intricate deformation and structural features under the ML method, which is somewhat time-consuming, laborious, and technically difficult to implement using traditional decay theory.

This letter consists of two parts. First, poor interpretability and opaque black-box algorithms are dilemmas that need to be urgently addressed in the development of machine learning. The quantum many-body atomic nuclear systems seem to provide a natural platform for the application of ML. Here we incorporate existing physical knowledge into the screening of ML inputs and perform a Pearson and Spearman correlation analysis \cite{correlation1,correlation2}, finding that the effective inputs selected in this way can achieve incredibly high accuracy in a single hidden layer neural network with only ten neurons. This simple structure, to some extent, increases the interpretability and weakens the black-box algorithm of ML.

Second, combining statistical ideas with the manifestation of ML in $\alpha$ decay theory, as expected, triggers more meaningful discoveries than the above-mentioned findings in ML itself. This work demonstrates the close connection between $P_\alpha$ and deformation from both statistical and numerical calculations, and further releases an unprecedented signal, that is, these atomic nuclei with hexadecanopole deformation are more likely to form $\alpha$ clusters. Two interesting phenomena accompany these quantitative studies: \textit{shape inheritance} and half-life \textit{inversion} due to shape staggering of adjacent even-even nuclei. Especially the significant difference in half-life caused by this inversion requires great attention in the identification of new nuclides. We also provide a method to analyze the relationship between half-life and shape from a dynamic perspective, i.e., treating the nucleus as a quasi-rigid body.

The remainder of the letter is organized as follows. First, correlation analysis is ongoing, and the Geiger-Nuttall law is perfectly demonstrated by the heat map, also the relationship between the decay energy and $P_\alpha$. Then we observe the shape inheritance phenomenon between the decaying parent nucleus and the daughter nucleus. Second, we systematically examine the performance of ML in $\alpha$ decay with the introduction of factors $P_\alpha$, $\beta_{2}$ and $\beta_{4}$ of the daughter and parent nucleus. In particular, for 41 super-heavy nuclides in the AME2020 mass table \cite{2020}, the precision of ML is 3.44 times higher than that of the improved Universal Decay Law (UDL) \cite{UDL}. Finally, we use the ML method to predict the half-life of the $Z=$119 and 120 isotope chains, both the WS4 \cite{WS4} and FRDM \cite{FRDM} databases as input, and observe this interesting inversion phenomenon not only in the unknown nuclear region but also in the region with experimental data.

\textit{Correlation analysis.}-
The ML currently popular in nuclear decay research is often relatively complex in structure, or a mixture of several ML methods is used, which to some extent complicates this mutual research and seems to have led to a fork in the road. In response to this situation, here we will try to screen effective inputs as a breakthrough.

We first clarify that the learning object of ML in this work is the logarithm of the experimental half-life ($\log_{10}T^\text{Exp.}$). Now review there are several physical knowledge or laws that are directly related to this quantity. Among them, there are two definite approximate linear relationships, namely the $\log_{10}T^\text{Exp.}$ and the square root of the ratio of the decay energy ($Q_\alpha^{-1/2}$) \cite{linear1}, the logarithm of the preformation factor ($\log_{10}P_{\alpha}$) and $Q_\alpha^{-1/2}$ \cite{linear2}. In addition, the deformation information, $\beta_{2}^\text{p,d}$ and $\beta_{4}^\text{p,d}$ of the decaying parent nuclei and the daughter nuclei from the WS4 mass table \cite{WS4} is incorporated directly into the input. 

\begin{figure}[htbp]
	\centering
	\includegraphics[width=0.5\textwidth]{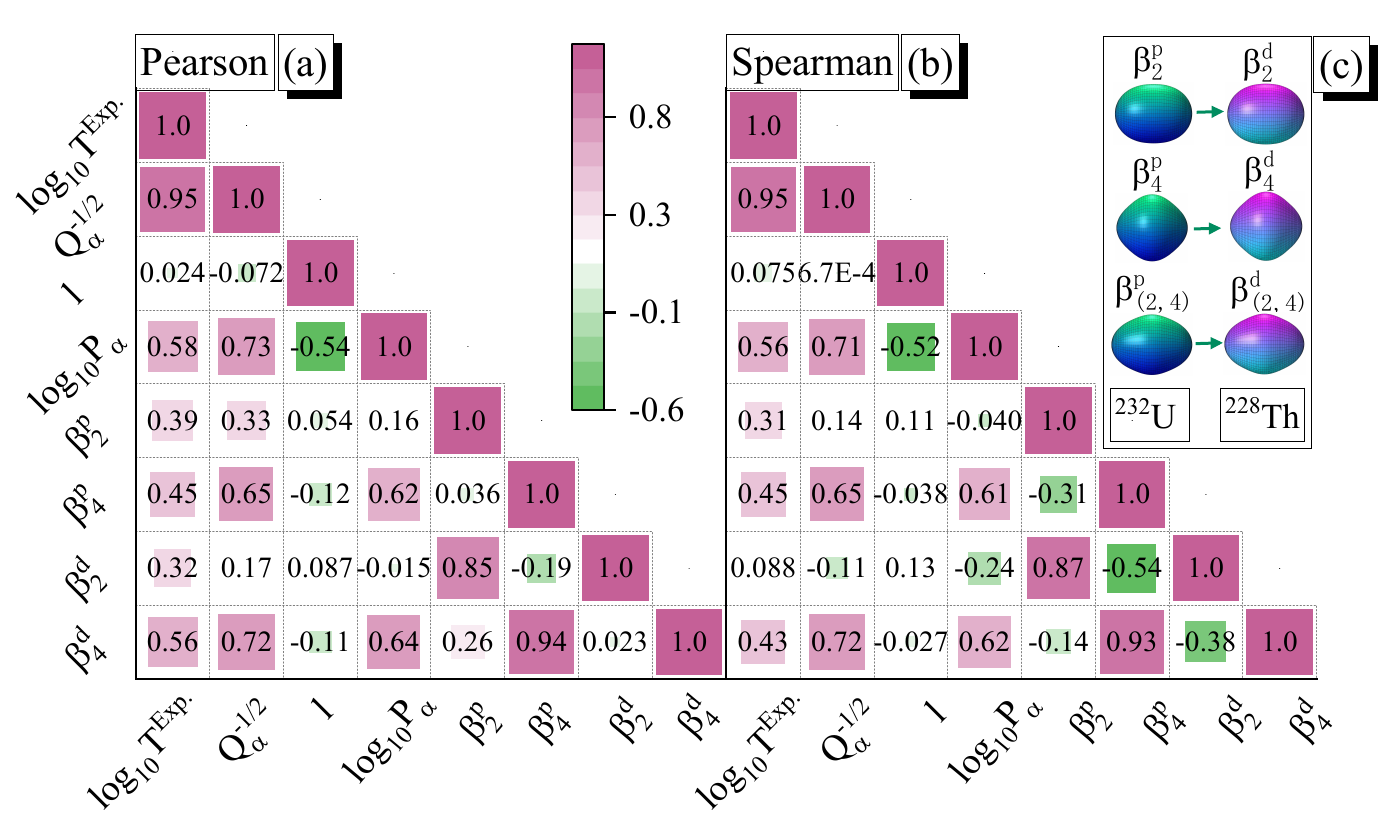}
	\caption{Multivariate correlations between half-life, decay energy, preformation factor, and deformation in nuclear decay dynamics: Pearson analyses (a) and Spearman analyses (b). Example of shape inheritance phenomenon (c), deformation data from WS4 \cite{WS4}.}\label{correlation}
\end{figure}

Pearson and Spearman correlations, two ideas originating from statistics \cite{correlation1,correlation2}, are used to advance the systematic analysis of correlations between the above-mentioned quantities. The corresponding heat maps are illustrated in Fig. \ref{correlation}. The results depict an extreme correlation between $\log_{10}T_\text{Exp.}$ and $Q_{\alpha}^\text{-1/2}$, which aligns closely with the Geiger-Nuttall law \cite{linear1}. Similarly, the close relationship between $Q_{\alpha}^\text{-1/2}$ and $\log_{10}P_{\alpha}$ also verifies the results in the Ref.\cite{linear2}. Remarkable correlations proclaim in the hexadecapole deformation parameters and decay energy, and preformation factors, which have never been found in previous studies, probably because the $\beta_{4}^\text{p,d}$ deformation is an order of magnitude smaller than the quadrupole deformation. Here, such a strong correlation may send the signal that nuclei with hexadecapole deformation are more likely to form $\alpha$ clusters, thereby promoting the occurrence of decay.

\textit{Shape inheritance.}-
The previous results strongly corroborate the reliability and robustness of our advent conclusion. This leads to an innovative concept: \textit{shape inheritance}, which means that the shapes of the parent and daughter nucleus in the decay system tend to remain consistent or similar. Fig. \ref{correlation} (c) takes the decay chain from the parent nucleus U$^{232}$ to the daughter nucleus Th$^{228}$ as an example. This phenomenon occurs frequently in the known nuclear region. This observation indicates that the structural properties of the parent nucleus are preserved to some extent in the daughter nucleus during $\alpha$ decay. Such a mechanism provides a novel perspective on the dynamic evolution of nuclear structures during the decay process, draws an intriguing parallel to the concept of inheritance in biological systems, offers a groundbreaking framework for understanding the continuity of nuclear deformation in the process of decay, and underscores the universality of generation continuity across seemingly disparate domains. 

This phenomenon can be traced back to previous studies. Refs.\cite{qichong1,Andreyev1,Andreyev2} indicate that a high degree of similarity of the parent and daughter nuclei leads to increased overlap in their wave functions. In other words, it is somewhat constrained by the $\beta_{2,4}$ deformation of the daughter and the parent nuclei. Hence, enhancing the probability of $\alpha$ decay. In addition, Ref. \cite{linear2} found that the closed shell effect of $Z=82, N=126$ is stronger than that of $Z=50, N=82$. From the analysis of the inheritance phenomenon, the shape of the parent nucleus $Z=82, N=126$ and its daughter nucleus $Z=80, N=124$ are quite different, that is, the overlap of the wave function is small, making it less likely to decay, resulting in a longer half-life and an obvious shell effect. Therefore, we provide a plausible explanation for the strength of this shell effect.

\textit{Machine learning performance.}-
We propose a UDL-driven ML approach that incorporates the UDL framework with a set of key physical parameters \cite{Na}, as the success of UDL \cite{UDL,Na}. In addition, the results of $\alpha$ decay can be easily compared after introducing the preformation factors and deformation. The proton ($Z$) and the mass number ($A$), $Q_{\alpha}^{-1/2}$ and the angular momentum ($l$) constitute the initial input, and after absorbing the UDL parameter ($x_\text{UDL}$), the input is $\{(Z,A,Q_{\alpha}^\text{-1/2},l); x_\text{UDL}\}$, which corresponds to
the UDL-driven ML method we mentioned. Successively absorbing $\beta_{2}^\text{p}, \beta_{4}^\text{p}$ of the parent nuclei and $\beta_{2}^\text{d}, \beta_{4}^\text{d}$ of the daughter nuclei, as well as $\log_{10}P_{\alpha}$. An interesting inspection is whether $\log_{10}T$ can be described well by a set of these effective quantities.

The datasets contain 369 known nuclei, 328 of which are used for training and 41 for testing. Here, a LeakyReLU is used as the activation function. The model is optimized using the Adam algorithm, with the root mean square (rms) deviation serving as the evaluation. Each set of different inputs was subjected to 10-fold cross-validation \cite{Pang,Na}, 100 times in total. The model whose rms value in both the training and test subsets is closest to the average of all iterations is selected as the representative. This approach ensures not only high performance, but also stability and reproducibility. 

\begin{figure}[htbp]
	\centering
	\includegraphics[width=0.5\textwidth]{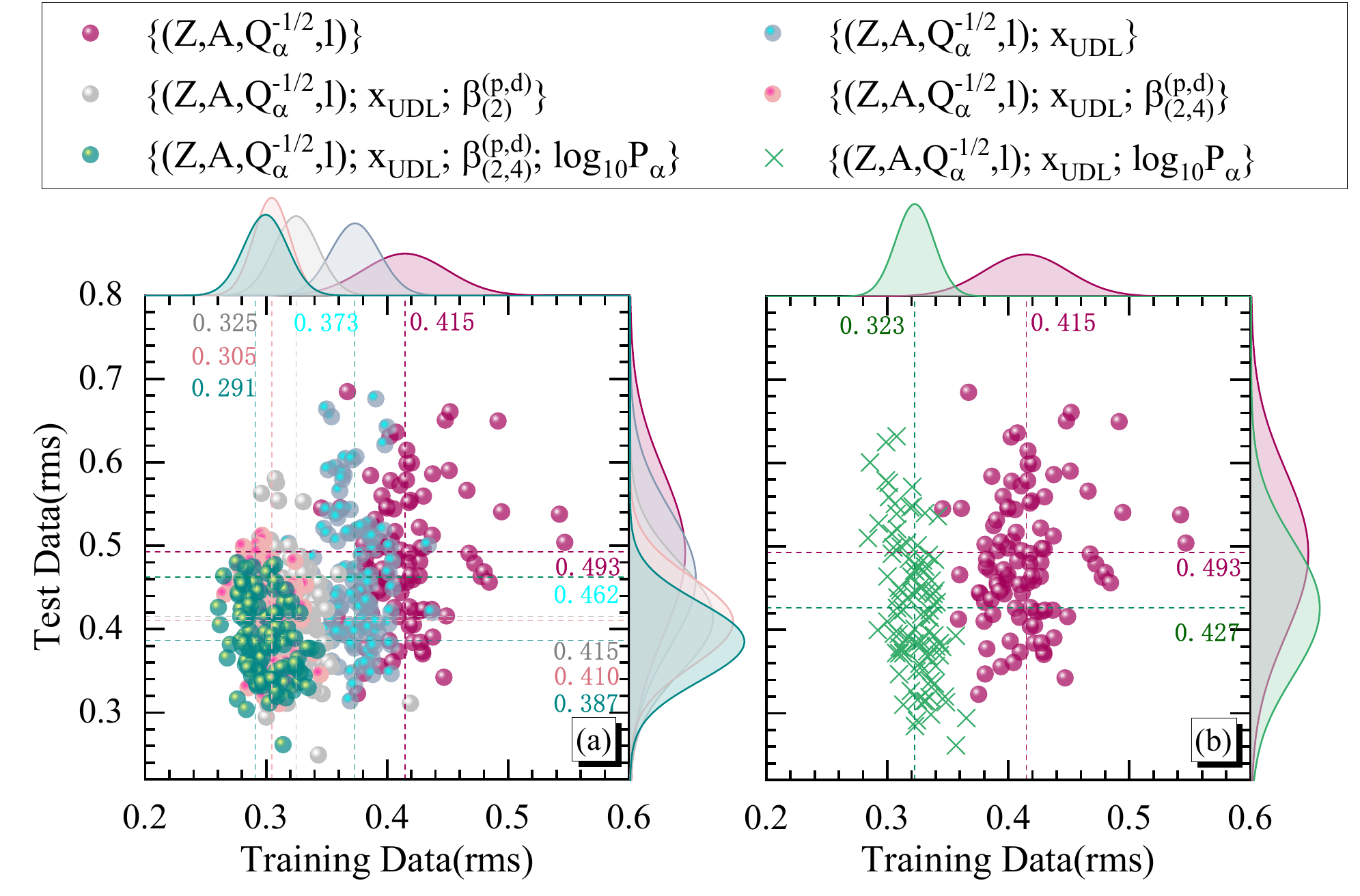}
	\caption{Performance of models across parameter groups: Distribution of rms values for training and testing datasets.}\label{100-rms}
\end{figure}

Fig. \ref{100-rms} plots the marginal statistical distribution of 100-times rms values for the training and test data, as well as the corresponding average rms for different inputs. With the stepwise inclusion of physical features, its accuracy has been significantly improved, rms on both training and test sets shows a clear downward trend. Especially compared to the theoretical accuracy of UDL, whose rms is 0.74. With the comprehensive parameter set \{($Z$, $A$, $Q_\alpha^\text{-1/2}$, $l$); $x_\text{UDL}$; $\beta_{2,4}^\text{p,d}$; $\log_{10}P_\alpha$\}, the average rms for 369 nuclei is 0.313. 

Note that the contribution of the formation amplitudes is significantly suppressed when the deformation is introduced into the ML first and then the preformation factor. This contrasts to Fig. \ref{100-rms}(b), where the contribution of the preformation factor alone is very significant. We can explain this by referring to the shape inheritance phenomenon mentioned above. The deformation changes of the daughter and parent nuclei of the same decay system in 369 nuclei are very small. This means that there is more overlap in wave functions, resulting in a higher probability of $\alpha$ cluster formation and a shorter half-life. Therefore, the contribution of $\log_{10}P_\alpha$ extracted by the empirical formula \cite{linear2} will become smaller, because this inheritance indirectly introduces the contribution of the preformation factor. From another perspective, the $P_\alpha$ given by the formula can be regarded as the \textit{residual} correction of the real $\alpha$ formation amplitudes in a sense. After all, factors such as the pairing effect \cite{preformation,pairing} will also affect the real $P_\alpha$. 

\begin{figure}[htbp]
	\centering
	\includegraphics[width=0.5\textwidth]{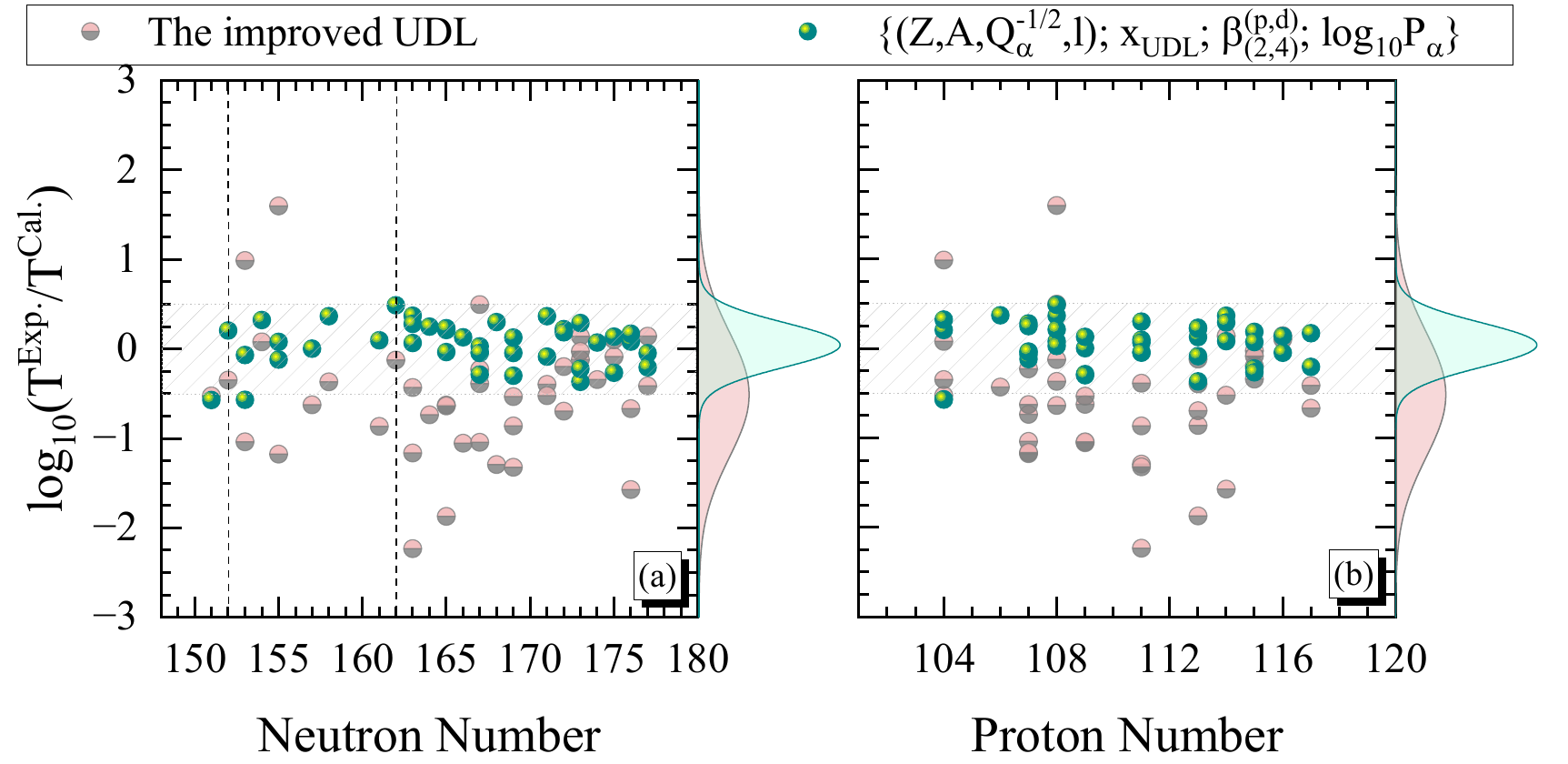}
	\caption{Evaluation of the prediction accuracy of $\alpha$ decay half-lives for 41 nuclei ($Z\geq104$): Distribution of differences between experimental values and ML, UDL calculations.}\label{41}
\end{figure}

ML has good generalization and predictive capabilities only when it performs well in a data set that has not participated in any training, that is, the test set. We predict the half-lives of 41 superheavy nuclei with $Z\geq104$ in AME2020 \cite{2020} using the ML method. Fig. \ref{41} shows marginal graphs of the difference between experimental and theoretical half-lives as a function of the number of neutrons (a) and the number of protons (b). The results of the improved UDL \cite{Na} were used for comparison. Specifically, our model achieves an rms of 0.26, which is significantly lower than 0.86 from the UDL \cite{Na}. The accuracy is significantly improved by 3.44 times. The sharp stark contrast between the UDL and our model is evident, with the differences predicted by the UDL being scattered. In contrast, our results are highly concentrated within $\log_{10}(T^\text{Exp.}/T^\text{Cal.})<0.5$, especially in the two closed subshell regions of $N = 152$ and 162, indicating that the ML method perfectly internalizes shell and sub-shell effect in the effective inputs, such as decay energies, preformation factors, and deformations. Therefore, effective input can avoid the direct introduction of a few magic numbers of protons and neutrons to increase the number of neurons, as in similar studies \cite{He}. Furthermore, we observe that the prediction errors become more consistent as we move towards heavier superheavy nuclei.

\begin{figure*}[htbp]
	\centering
	\includegraphics[width=1\textwidth]{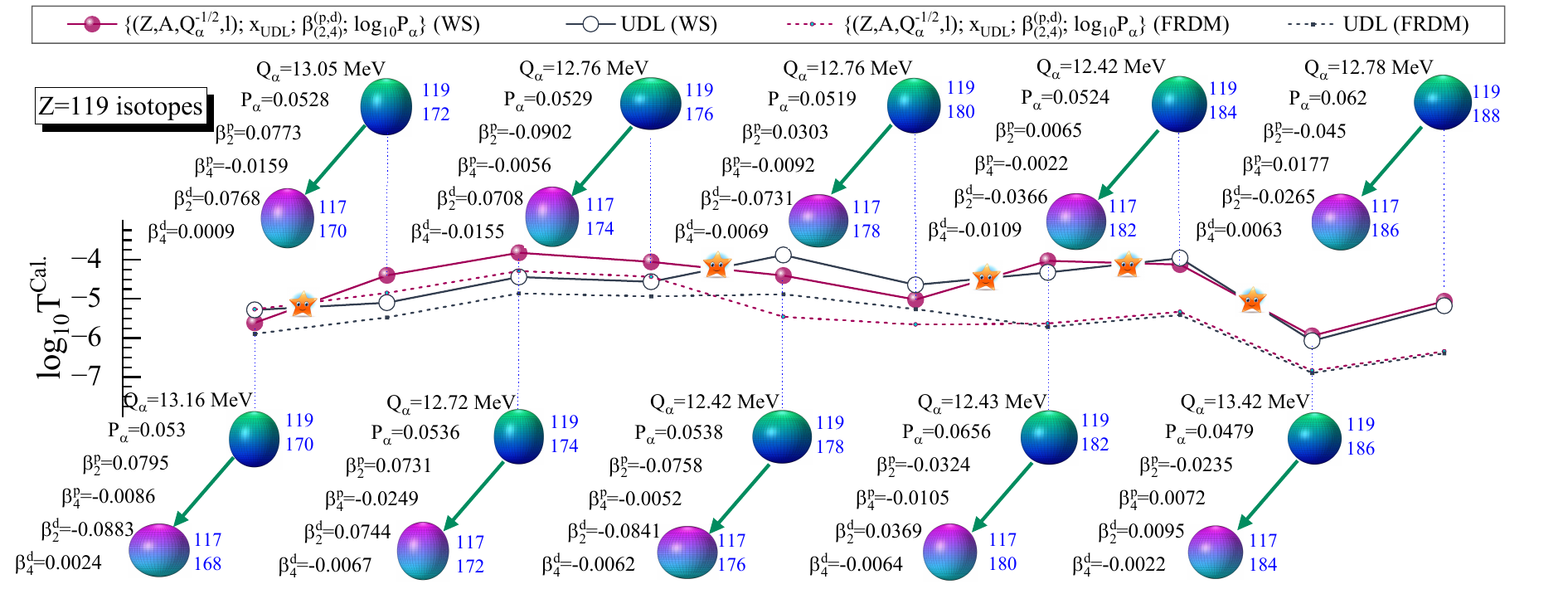}
	\caption{For even $Z=119$ isotopes: Prediction results of ML vs UDL. The shape parameters and decay energy data of the parent and daughter nuclei in the same decay system are extracted from the WS4 database. The preformation factor is extracted from the empirical formulac \cite{linear2} with WS4 decay energy as input. The red balls and empty circles are the prediction results given by the WS4 database as input. The red and the gray dotted lines correspond to the data results of FRDM.}\label{SHN-isotopes}
\end{figure*}

\textit{Inversion.}-
We extend predictions for elements $Z=119$ and $Z=120$ using decay energies and deformations of the WS4 and the FRDM. Fig. \ref{SHN-isotopes} shows the predicted half-life for the even 119 isotope chain, as well as the shapes of each parent nucleus and its daughter, with $\beta_2$ and $\beta_4$ fused together, here using the shape given by the WS4 data as an example. The information of the even 120 isotope chain is listed in Table \ref{120}, including both WS4 and FRDM.

There is an interesting pattern hidden in Fig. \ref{SHN-isotopes}.
For daughter nuclei with an oblate configuration, the ML predictions tend to be shorter than the UDL predictions. In contrast, daughter nuclei with prolate deformation exhibit longer half-lives. We conduct a brief analysis of this phenomenon from a dynamics perspective and find that, compared with the prolate ellipsoid, the oblate ellipsoid has a small difference in main moment of inertia and high rotation stability. So, if the daughter nucleus is oblate and the parent nucleus is prolate, naturally a system will evolve from an unstable state to a stable state. Therefore, this pattern matching will promote the occurrence of $\alpha$ decay, such as the decay system $^{289}$119 to $^{285}$Ts. If it is the opposite matching pattern, it will hinder the occurrence of decay (see $^{295}$119 to $^{291}$Ts). 

\begin{figure}[htbp]
	\centering
	\includegraphics[width=0.47\textwidth]{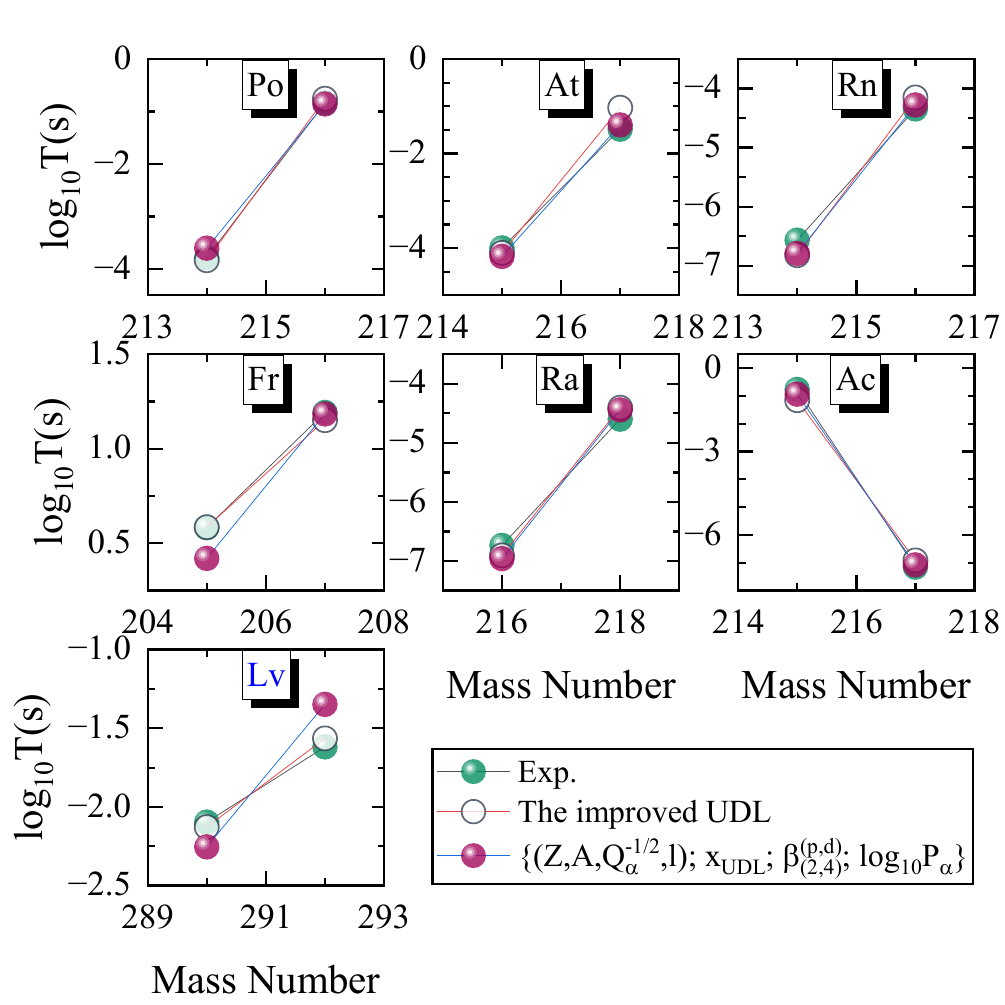}
	\caption{The inversion phenomenon present in 369 known nuclear regions is caused by changes in the $\beta_2$ deformation configuration of the parent and daughter nuclei.}\label{inversion}
\end{figure}

\begin{table}[htbp]
	\caption{The quadrupole deformation and half-life given by ML and UDL for even $Z=120$ isotope chain. The inversion event caused by the $\beta_2$ deformation of the adjacent nucleus also exists.}
	\begin{center}\begin{tabular}{c||cccc|cccc}
			\hline\hline
			\multirow{2}*{Parent} & \multicolumn{4}{c}{WS4}& \multicolumn{4}{c}{FRDM}\\\cline{2-9}
			&$\beta_2^{p}$&$\beta_2^{d}$&ML&UDL&$\beta_2^{p}$&$\beta_2^{d}$&ML&UDL\\\hline
			$^{290}$120 &0.081&-0.095 &-6.85&-6.82  &-0.122&0.08 &-5.85&-6.90\\
			$^{292}$120 &0.079&0.078&-5.36&-6.40    &-0.130&0.08 &-5.91&-6.97\\
			$^{294}$120&0.076&0.075 &-5.02&-5.99    &0.081&-0.112 &-6.72&-6.47\\
			$^{296}$120&0.076&0.072 &-5.31&-6.22    &-0.096&0.081 &-5.76&-6.70\\
			$^{298}$120&-0.075&-0.09&-5.75&-5.58    &-0.079&-0.087 &-6.25&-6.04\\
			$^{300}$120&0.022&-0.077&-6.26&-6.24    &-0.008&-0.079 &-7.03&-6.95\\
			$^{302}$120&-0.028&0.038&-4.77&-5.40    &0.00&-0.008 &-6.37&-6.74\\
			$^{304}$120&-0.001&-0.038&-4.97&-5.16   &0.00&0.00 &-6.29&-6.75\\
			$^{306}$120&-0.020&0.012&-6.71&-7.24    &0.00&0.00 &-7.69&-8.12\\
			$^{308}$120&-0.045&-0.025&-5.03&-5.65   &0.001&0.00 &-5.18&-5.65\\\hline
			\hline
		\end{tabular}\label{120}
	\end{center}
\end{table}

This pattern inspires another interesting finding in Fig. \ref{SHN-isotopes}, which we term half-life \textit{inversion}, a phenomenon caused by the shape staggering of adjacent even-even nuclei. 
Specifically, differences in nuclear shape are always observed for adjacent even-even nuclei at both ends of the intersection of the ML-predicted curve and the UDL curve, whether in the parent nucleus, the daughter nucleus, or both, suggesting a potential link between shape transitions and changes in decay behavior. Note that this relationship is not reciprocal, as not all differences in nuclear shape correspond to such a crossover. It is worth mentioning that the results given by FRDM also show an inversion phenomenon. This phenomenon is obvious in the isotope chains of 119 and 120, due to the frequent shape staggering of the daughter nuclei and the parent nuclei. For the $^{292}$120 nuclei in Table \ref{120}, the predicted half-lives ($T^\text{Pre.}$) given by ML and UDL differ by a factor of ten. This apparent discrepancy must be taken seriously in experiments because of the dramatic shape changes. 

This inversion phenomenon actually exists in the 369 nuclei, but in these known nuclear regions, the $\beta_{2,4}$ values of the daughter nuclei in two adjacent decay systems are very close to those of the parent nucleus. Therefore, previous studies did not find this inversion effect. We now reexamine the 369 nucleus and find that this inversion effect also occurs in decay systems where the shapes of adjacent even-even nuclei change significantly. Fig. \ref{inversion} lists seven obvious inversion cases. Among them, Po, At, Rn, Fr, Ra, and Ac are basically concentrated in the area near the neutron-deficient Pb nucleus. Previous experiments \cite{Heyde,Andreyev3} have found that this region is rich in shape coexistence, with nuclei exhibiting a variety of shape configurations, such as spherical, oblate and prolate, and rapidly transitioning between them, which is consistent with why this region frequently experiences such reversal events. In addition, compared to the quadrupole deformation, the change rate of the hexadecapole deformation ($\beta_4$) is more obvious, which will also cause a smaller inversion, although its value is one order of magnitude smaller than $\beta_2$. We find more than twenty such inversion events. However, because of the inevitable systematic errors of the UDL, it is not conducive to strictly distinguish whether the inversion is caused by the change of the hexadecapole deformation or by the error.

By integrating the inheritance phenomenon caused by shape maintenance and the inversion phenomenon caused by shape change in the same decay system, the deformation of the atomic nucleus must take into account the deformation configuration of both the daughter nucleus and the parent nucleus, and at the same time include quadrupole and hexadecapole deformations, which have a deep intrinsic connection with the decay theory, especially the kinetic process of the decay theory.

\textit{Conclusions.}-This letter uses the knowledge and laws of nuclear decay and statistical correlation analysis to achieve a high accuracy of 0.313 rms on 369 known nuclei using a single hidden layer with only 10 neurons, and improves the RMS accuracy by 3.44 times on 41 test nuclei. This approach to simplifying machine learning architecture provides a new paradigm for developing physics-driven neural networks. 

From the perspective of decay physics, this study proves the Geiger-Nuttall law from a statistical point of view, confirms the linear relationship between the decay energy $Q_\alpha^{-1/2}$ and the logarithm of the preformation factor $\log_{10}P_\alpha$ and extract a signal that nuclei with hexadecapole deformation are more likely to produce $\alpha$ clusters. The most important finding of this work is the link between shape inheritance and shape staggering effect and half-life. The deformation configuration of the parent nucleus before decay and the configuration of the emitted daughter nucleus simultaneously reflect the dynamic process of the decay theory. It is incomplete to consider only the shape of the parent nucleus or the daughter nucleus. This provides a strong signal for the experimental synthesis of elements in the eighth period of the periodic table, $Z$ = 119 and 120. Observing the shapes of the daughter nuclei, Ts ($Z$ = 117) and Og ($Z$ = 118), one can speculate that the half-life is longer or shorter than predicted by traditional theoretical decay models.

\begin{acknowledgments}

  The authors thank Jing Geng for insightful comments and Professor Jarah Evslin for polishing the English of the article. This work is supported by the National Natural Science Foundation of China (Grants No. 12105128 and No. 12175170) and the National Key Research and Development (R\&D) Program under Grant No. 2021YFA1601500.
\end{acknowledgments}

\end{document}